\documentclass[aps,twocolumn,groupedaddress,showpacs,prb,preprintnumbers,raggedbottom]{revtex4}
\usepackage{graphicx}

\newcommand{\cecoin}    {CeCoIn$_5$}

\newcommand{\cerhin}     {CeRhIn$_5$}

\begin{document}

\preprint{LA-UR-03-8206}
\title{Scaling in the Emergent Behavior of Heavy Electron Materials}
\author{N. J. Curro}
\author{B.-L. Young}
\affiliation{Condensed Matter and Thermal Physics, Los Alamos National Laboratory, Los
Alamos, NM 87545, USA}
\author{J. Schmalian}
\affiliation{Department of Physics and Astronomy and Ames Laboratory, Iowa State
University, Ames, Iowa 50011, USA}
\author{D. Pines}
\affiliation{Institute for Complex Adaptive Matter, University of California and\\
Theoretical Division, Los Alamos National Laboratory, Los Alamos, NM 87545,
and Department of Physics, University of Illinois at Urbana-Champaign,
Urbana, IL, 61801, USA}
\date{\today}

\begin{abstract}
We show that the NMR Knight shift anomaly exhibited by a large number of
heavy electron materials can be understood in terms of the different
hyperfine couplings of probe nuclei to localized spins and to conduction
electrons. The onset of the anomaly is at a temperature $T^*$, below which
an itinerant component of the magnetic susceptibility develops. This second
component characterizes the polarization of the conduction electrons by the
local moments and is a signature of the emerging heavy electron state. The
heavy electron component grows as $\log T$ below $T^*$, and scales
universally for all measured Ce, Yb and U based materials. Our results
suggest that $T^*$ is not related to the single ion Kondo temperature, $T_K$%
, but rather represents a \textit{correlated} Kondo temperature that
provides a measure of the strength of the intersite coupling between the
local moments. Our analysis strongly supports the two-fluid description of
heavy electron materials developed by Nakatsuji, Pines and Fisk.
\end{abstract}

\pacs{71.27.+a,75.20.Hr,76.60.Cq}
\maketitle

\thispagestyle{myheadings} \markright{{\em LA-UR-03-8206}}

\subsection{Introduction}

The Kondo lattice is a paradigm for heavy electron materials.
Recently a number of puzzling experimental observations  have been
made in systems close to a magnetic phase transition at low
temperatures,\cite{Custers,Gegenward,Steward,HvL,Schroder} and
several new theoretical approaches for non-mean field quantum critical behavior have been proposed%
\cite{Si,Piers,Piers2,Lavagna,Sachdev}. A complete theoretical
description of the Kondo lattice remains elusive after two decades
since the discovery of heavy electron behavior. However,  several
common experimental signatures have been identified in these
materials that must be captured by any theoretical description. In
particular, heavy fermion and mixed valent systems exhibit a
crossover between localized moments at high temperatures to
coherent behavior at low temperatures. Typically this crossover is
evident as a broad maximum in the resistivity; in some cases the
bulk magnetic susceptibility also exhibits a maximum, although not
always at the same temperature as the resistivity. This behavior
has traditionally been understood as the onset of coherent
scattering of conduction electrons by the Kondo lattice of 4f
spins: at high temperatures ($T>T_{\mathrm{coh}}$) the 4f spins
scatter the
conduction electrons as independent local impurities; however, below $T_{%
\mathrm{coh}}$ and at low temperatures the Kondo lattice behaves in a
coherent fashion. Although a microscopic theory of this process has not
emerged, experimental signatures of this crossover are clearly evident in
many Ce, Yb and U based compounds.

\begin{figure}[tbp]
\centering  \includegraphics[width=\linewidth]{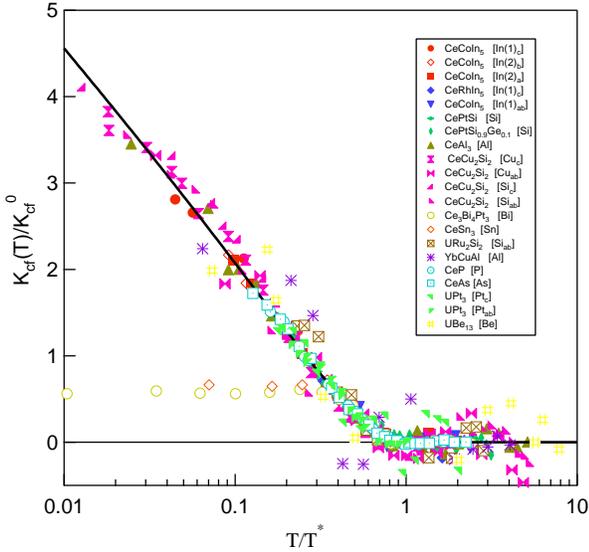}
\caption{$K_\mathrm{cf}(T)/K_\mathrm{cf}^0$ versus $\ln(T/T^*)$ for several
Kondo lattice systems, showing the scaling behavior of the Kondo liquid
component of susceptibility. The solid line is given by Eq. (\protect\ref%
{eqn:KKL}).}
\label{fig:scalingplot}
\end{figure}

In the majority of heavy electron and mixed valent materials for which
Knight shift measurements exist, it has consistently been observed that
below a temperature $T^*$, the NMR (as well as $\mu$SR) Knight shift, $K$,
fails to track the bulk susceptibility, $\chi$.\cite%
{narath,maclaughlinreview} The reason for this anomalous behavior has
remained elusive. The Knight shift measures the field at the nucleus brought
about by the hyperfine interaction with the electrons. When the electrons
are polarized in an external magnetic field, they create a hyperfine field
at the nuclei that is proportional to $\chi$. If there is only one magnetic
component, then $K\sim\chi$.

Traditionally the breakdown of this relationship has been attributed to
local phenomena associated with the 4f electrons. In the crystal field
scenario, the hyperfine coupling changes when the excited states of the
crystal field split 4f electron become depopulated. \cite{yasuokaCeCu2Si2}
In the Kondo impurity scenario, $T^*$ is the Kondo temperature, below which
the 4f electrons are screened by the conduction electrons and the bulk $\chi$
is reduced.\cite{cox} The Knight shift measures the local susceptibility,
which is not screened; and therefore the linear relationship between $K$ and
$\chi$ breaks down. Still other authors have explained this anomaly in terms
of a temperature dependent hyperfine coupling that is modified by the onset
of coherence.\cite{sonier}

Here we propose that the origin of this anomaly is collective rather than
local, and that $T^{\ast }$ is the temperature at which the heavy electron
liquid begins to emerge from the Kondo lattice of localized 4f spins. \ We
demonstrate that below $T^{\ast }$ the polarization of the background
conduction electron spin system by the correlated $f$-spins is characterized
by a distinct and universal temperature dependence. This polarization is
characterized by the magnetic susceptibility, $\chi_\mathrm{cf} = \langle
\mathbf{S}_c \mathbf{S}_f \rangle$, where $\mathbf{S}_c$ and $\mathbf{S}_f$
are the conduction and local moment spins, respectively. We show that the
two-fluid description of the Kondo lattice proposed by Nakatsuji, Pines and
Fisk (NPF) provides a quantitative explanation for this anomalous behavior.%
\cite{NPF} In turn our analysis allows us to give a more microscopic
interpretation for the two fluids introduced by NPF. Our results are
consistent with NPF who argued that $T^{\ast }$ is a \emph{correlated} Kondo
temperature, strongly affected by the intersite $f$-electron interaction,
rather than the familiar single-ion Kondo temperature, $T_{K}$. \ In
addition, we are able to determine quantitatively the temperature evolution
of the heavy electron spin susceptibility by combining measurements of the
temperature dependence of the Knight shift with those of the bulk magnetic
susceptibility. \ We find that an excellent fit to existing experimental
data in fourteen\ heavy electron and mixed valent systems is obtained with a
susceptibility whose temperature dependence follows the simple form:
\begin{equation}
\chi _\mathrm{cf}\sim \left( 1-\frac{T}{T^{\ast }}\right) \log \frac{T^{\ast
}}{T},  \label{eqn:chiHE}
\end{equation}%
a result that suggests that in a Kondo lattice the emergent behavior of the
heavy electron liquid can be characterized quite generally by the single
energy scale, $T^{\ast }$, that NPF have proposed is a direct measure of the
strength of nearest neighbor intersite magnetic coupling. For CeSn$_{3}$ and
Ce$_{3}$Bi$_{4}$Pt$_{3}$ we find quite similar behavior, except that below a
cut-off temperature, $T_{0}$, the Knight shift once more tracks the bulk
susceptibility. We argue that below $T_{0}$ the formation of the heavy
electron liquid is complete, and the system has a single, itinerant magnetic
component.

\begin{figure}[tbp]
\centering \includegraphics[width=\linewidth]{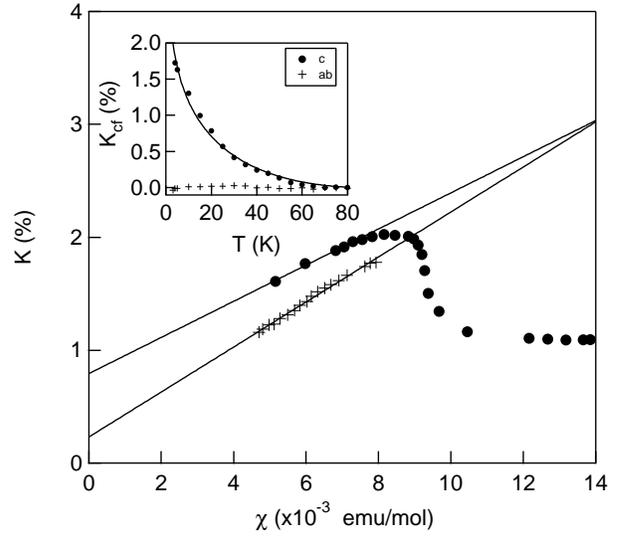}
\caption{The In(1) Knight shift in CeCoIn$_5$\ versus the bulk
susceptibility.\protect\cite{curroCeCoIn5} The solid lines are fits to the
high temperature data. Inset: $K_\mathrm{cf}$ versus $T$, and a fit to Eq. (
\protect\ref{eqn:KKL}).}
\label{fig:In1CeCoIn5}
\end{figure}

In Section B, we discuss the origin of the Knight shift in a Kondo lattice,
and its anomalous behavior below $T^*$ in the two-fluid description of NPF.
We present the experimental data on fourteen\ Kondo lattice materials, and
show that the simple expression given by Eq. (\ref{eqn:chiHE}) provides a
quantitative account of the existing results. For the heavy electron
material, CeCoIn$_5$, the second component of susceptibility that we obtain
by Knight shift measurements is shown to be in excellent agreement with that
deduced by NPF. We present our discussion of these results and our
conclusions in Section C, and give the data used to obtain our results in
the Appendix.

\subsection{Knight Shifts in Kondo Lattice}

In the two fluid description proposed by NPF to explain bulk specific heat
and susceptibility measurements in La doped CeCoIn$_{5}$, the authors
postulate that a fraction $f(T)$ of the 4f electrons in the Kondo lattice
become delocalized below $T^{\ast }$, forming a coherent state, the heavy
electron liquid, analogous to the superfluid component of $^{4}$He. $f(T)$
resembles an order parameter for the coherent heavy electron component,
while the fraction of the 4f (5f) electrons remaining localized resembles
the normal fluid component. The magnetic system contains one component that
is localized on the magnetic sites $i$ with susceptibility $\chi _{\mathrm{KI%
}}$, and a second component that is associated with the itinerant heavy
quasiparticles with susceptibility $\chi _{\mathrm{HF}}$:
\begin{equation}
\chi (T)=[1-f(T)]\chi _{\mathrm{KI}}(T)+f(T)\chi _{\mathrm{HF}}(T).
\label{eqn:chitot}
\end{equation}%
Detailed insight into the emergence of two contributions to the
susceptibility with dramatically different $T$-dependency can be obtained by
taking into account that the total spin of the system is the sum of the
localized $f$-electron and the conduction electron spins%
\begin{equation}
\mathbf{S}_{\mathrm{tot}}=\sum_{i}\mathbf{S}^{\mathrm{f}}\left( \mathbf{r}%
_{i}\right) +\sum_{l}\mathbf{S}^{\mathrm{c}}\left(
\mathbf{r}_{l}\right)
\label{eqn:stot}
\end{equation}%
where $\mathbf{r}_{i}$ are the positions of the $f$-electrons and $\mathbf{r}%
_{l}$ the positions of the itinerant component of the system within a
Wannier orbital representation. $\mathbf{S}^{\mathrm{c}}\left( \mathbf{r}%
_{l}\right) =\frac{1}{2}\sum_{\sigma \sigma ^{\prime }}c_{\mathbf{r}%
_{l}\sigma }^{\dagger }\mathbf{\sigma }_{\sigma \sigma ^{\prime }}c_{\mathbf{%
r}_{l}\sigma ^{\prime }}$ is the conduction electron spin density. Our
results show that at $T^{\ast }$, as a result of the coupling to the $f$%
-electron spins, the conduction electrons acquire a heavy component,
characterized by the correlation function $\left\langle \mathbf{S}^{\mathrm{f%
}}\left( \mathbf{r}_{i}\right) \mathbf{S}^{\mathrm{c}}\left( \mathbf{r}%
_{l}\right) \right\rangle $. The uniform susceptibility is now given as $%
\chi =\frac{1}{N}\frac{\partial }{\partial H}\left\langle \mathbf{M}_{%
\mathrm{tot}}\right\rangle $ where $\mathbf{M}_{\mathrm{tot}}=\sum_{i}%
\mathbf{S}^{\mathrm{f}}\left( \mathbf{r}_{i}\right) +\sum_{l}\mathbf{S}^{%
\mathrm{c}}\left( \mathbf{r}_{l}\right) $ and it follows that%
\begin{eqnarray}
\chi &=&\chi _{\mathrm{\mathrm{ff}}}+2\chi _{\mathrm{cf}}+\chi _{\mathrm{cc}}
\label{eqn:susorb} \\
&\approx &\chi _{\mathrm{\mathrm{ff}}}+2\chi _{\mathrm{cf}},  \nonumber
\end{eqnarray}%
where $\chi _{\mathrm{\mathrm{ff}}}=\ \frac{1}{N}\sum_{i,i^{\prime
}}\left\langle \mathbf{S}^{\mathrm{f}}\left( \mathbf{r}_{i}\right) \mathbf{S}%
^{\mathrm{f}}\left( \mathbf{r}_{i^{\prime }}\right) \right\rangle $ and $%
\chi _{\mathrm{cf}}=\frac{1}{N}\sum_{i,l}\left\langle \mathbf{S}^{\mathrm{f}%
}\left( \mathbf{r}_{i}\right) \mathbf{S}^{\mathrm{c}}\left( \mathbf{r}%
_{l}\right) \right\rangle $ are the orbital resolved susceptibilities,
characterizing the magnetic response of the pure $\mathrm{f}$-system as well
as the polarization of the background conduction electron spin system by the
correlated $\mathrm{f}$-spins, respectively. We recover the NPF result, Eq. (%
\ref{eqn:chitot}), by neglecting $\chi _{\mathrm{cc}}$, the uniform
susceptibility of the background conduction electrons, which is small, and
identifying $\chi _{\mathrm{ff}}=[1-f(T)]\chi _{\mathrm{KI}}$ and $2\chi _{%
\mathrm{cf}}=f(T)\chi _{\mathrm{HF}}$.

Quite generally, we expect the hyperfine couplings associated with these two
magnetic components to differ. For the local moments, the dominant hyperfine
interaction is via a transferred coupling between the nuclei (typically at a
different crystalline site) and the local moment. A finite spin density is
induced on the neighboring nucleus via wavefunction overlap, or via an
indirect interaction mediated by conduction electrons. In general, the
transferred hyperfine interaction may couple the nucleus to several nearest
neighbor local moment sites. On the other hand, if the magnetic component is
delocalized as in a Fermi liquid, it has an additional on-site hyperfine
contact term, which generally dominates. For a Kondo lattice system that
retains aspects of both localized and delocalized behavior, one can
reasonably expect both contact as well as transferred hyperfine couplings. A
similar situation is found in the cuprate superconductors.\cite{milarice}

We therefore postulate the following hyperfine Hamiltonian:
\begin{equation}
\mathcal{H}_{\mathrm{hyp}}=\gamma \hbar \sum_{l}\mathbf{I}\left( \mathbf{r}%
_{l}\right) \mathbf{\cdot A\cdot S}^{\mathrm{c}}\left( \mathbf{r}_{l}\right)
+\gamma \hbar \sum_{i,l}\mathbf{I}\left( \mathbf{r}_{l}\right) \mathbf{\cdot
B}_{i}\mathbf{\cdot S}^{\mathrm{f}}\left( \mathbf{r}_{i}\right) ,
\label{eqn:hyperfine}
\end{equation}%
where $\mathbf{A}$ and $\mathbf{B}$ are the temperature independent contact
and transferred hyperfine tensors, respectively, and $\mathbf{r}_{i}$ are
positions of the nearest neighbor 4f (5f) sites. The Knight shift is given
by: $\mathcal{H}_{\mathrm{hyp}}=\gamma \hbar \sum_{l}\mathbf{I}\left(
\mathbf{r}_{l}\right) \cdot \mathbf{K}\cdot \mathbf{H}_{0}$, where $\mathbf{H%
}_{0}$ is the applied field. By recognizing that $\langle \mathbf{S}^{%
\mathrm{c}}\left( \mathbf{r}\right) \rangle =\chi _{\mathrm{cf}}\mathbf{H}%
_{0}$ and $\langle \mathbf{S}^{\mathrm{f}}\left( \mathbf{r}\right) \rangle
=\left( \chi _{\mathrm{cf}}+\chi _{\mathrm{\mathrm{ff}}}\right) \mathbf{H}%
_{0}$, and making use of Eq. (\ref{eqn:hyperfine}), we then have:
\begin{equation}
K_{\alpha }\left( T\right) =K_{0,\alpha }+\left( A_{\alpha }+B_{\alpha
}\right) \chi _{\mathrm{cf}}\left( T\right) +B_{\alpha }\chi _{\mathrm{ff}%
}\left( T\right)
\end{equation}%
where $K_{0,\alpha }$ is an offset (to account for orbital susceptibility
and other $T$ independent effects), and we have dropped the summation over
the neighboring sites for simplicity and incorporated the couplings into the
constant $B$. For $T>T^{\ast }$, $\ $\ where the linear relationship between
Knight shift and bulk susceptibility holds, we make the assumption that $%
\chi _{\mathrm{cf}}(T)\simeq 0$. This allows us to determine the coupling
constant $B$. \ In the dilute limit where the single impurity Kondo problem
applies, one can carry out an explicit calculation, where indeed one finds
for large temperatures ($T\gg T_{\mathrm{K}}$) $\frac{\chi _{\mathrm{cf}}}{%
\chi _{\mathrm{ff}}}\simeq \rho _{F}J_{K}\ll 1$. \cite{BarzykinAffleck}
Below $T^{\ast }$, $\chi (T)$ and $K_{\alpha }\left( T\right) $ are no
longer proportional. $\chi _{\mathrm{cf}}(T)$ and $\chi _{\mathrm{ff}}\left(
T\right) $ enter into $\chi (T)$ and $K_{\alpha }\left( T\right) $ with
different weights, which is due to the additional hyperfine coupling
constants in the Knight shift. This allows us to separate the two
contributions to the susceptibility from a knowledge of bulk susceptibility
and Knight shift. In particular, we obtain the crucial relationship%
\begin{eqnarray}
K_{\mathrm{cf},\alpha }(T) &=&K_{\alpha }(T)-K_{0,\alpha }-B_{\alpha }\chi
(T)  \nonumber \\
&=&\left( A_{\alpha }-B_{\alpha }\right) \chi _{\mathrm{cf}}(T).
\label{eqn:chiKL}
\end{eqnarray}%
This enables us to single out the heavy electron component, $\chi _{\mathrm{%
cf}}$, that must be thought of as a hybridized many body state where the
delocalized nature of the $f$-spin degrees of freedom is made explicit. In
Eqs. \ref{eqn:stot} and \ref{eqn:susorb} we made the simplifying assumption that the $g$%
-factors of the localized and conduction electron spins are the
same. Including different $g$-factors for the two spins will not
change the relation $K_{\mathrm{cf},\alpha }(T)\propto \chi
_{\mathrm{cf}}(T)$ but only affect the numerical value of unknown
prefactor $A_{\alpha }-B_{\alpha }\rightarrow A_{\alpha
}-\frac{g_{\mathrm{f}}}{g_{\mathrm{c}}}B_{\alpha }$. We note that
a necessary condition for the existence of a Knight shift anomaly
is that $A_{\alpha }\neq B_{\alpha }$. In Figs.
\ref{fig:In1CeCoIn5}
and \ref{fig:In2CeCoIn5} we show this anomaly in CeCoIn$_{5}$. $K_{0}$ and $%
B_{\alpha }$ are determined by fitting the high temperature data, shown as
the solid lines; these values are given in Table \ref{tbl:hyperfine}. Below $%
T^{\ast }$ we obtain the temperature dependence of the heavy electron
component, $K_{\mathrm{cf},\alpha }(T)$, which is shown in the inset. Note
that without an independent measure of $\chi _{\mathrm{cf}}(T)$, the on-site
coupling $\mathbf{A}$ remains undetermined.

In Fig. \ref{fig:scalingplot} we plot $K_{\mathrm{cf},\alpha }(T)$ versus $%
T/T^{\ast }$ for fourteen\ heavy electron and mixed valent systems. $T^{\ast
}$ is experimentally determined as the temperature below which $\chi (T)$
and $K_{\alpha }(T)$ cease to be proportional to each other. The collapse of
data for such a considerable number of systems is particularly impressing
and is the single most important observation of this paper. This universal
behavior of $\chi _\mathrm{cf}$ is particularly surprising if one takes into
account that the bulk susceptibility $\chi $ and the total shift $K_{\alpha
}(T)$ behave qualitatively differently for a number of the compounds shown.
It is $\chi _{\mathrm{cf}}$ which is universal for all these materials.

\begin{table*}[tbp]
\caption{The Knight shift parameters in several Kondo lattice systems.}
\label{tbl:hyperfine}%
\begin{ruledtabular}
\begin{tabular}{lccccclcl}
  Material (site)$^{\rm Ref.}$ & $T^* (K) $ & $K_0(\%)$ & $B_{\alpha}$ (kOe/$\mu_{\rm B}$)  & $A_{\alpha}$ (kOe/$\mu_{\rm B}$) & $K_\mathrm{cf}^0 (\%)$ &  $\gamma$ (mJ/mol K$^2$)$^{\rm Ref.}$\\
  \hline
  \cecoin\ (In(1)$_c$)\cite{curroCeCoIn5} & 89 & 0.79 & 8.9 & 13.7 & 3.3 & 290\cite{CeCoIn5ref}\\
  \cecoin\ (In(1)$_{ab}$)\cite{curroCeCoIn5} & - & 0.13 & 12.1 & 12.1 & - & 290\cite{CeCoIn5ref}\\
  \cecoin\ (In(2)$_a$)\cite{curroCeCoIn5} & 42 & 1.14 & -0.4 & -5.9& -2.0  &  290\cite{CeCoIn5ref}\\
  \cecoin\ (In(2)$_b$)\cite{curroCeCoIn5} & 42 & 0.77 & 10.3  & -4.1& -1.3 &  290\cite{CeCoIn5ref}\\
  \cecoin\ (In(2)$_c$)\cite{curroCeCoIn5} & 95 & -2.43 & 28.1 & 12.1& 3.1  &  290\cite{CeCoIn5ref}\\
  CeCu$_2$Si$_2$ (Cu$_{c}$)\cite{yasuokaCeCu2Si2} & 171  & 0.04 & -0.2 & - & -0.3 & 700\cite{CeCu2Si2ref} \\
  CeCu$_2$Si$_2$ (Cu$_{ab}$)\cite{yasuokaCeCu2Si2} & 58  & -0.05 & 2.5 & - & -0.1 & 700\cite{CeCu2Si2ref}\\
  CeCu$_2$Si$_2$ (Si$_{c}$)\cite{yasuokaCeCu2Si2} & 171  & 0.12 & 2.7 & - & -0.3 & 700\cite{CeCu2Si2ref} \\
  CeCu$_2$Si$_2$ (Si$_{ab}$)\cite{yasuokaCeCu2Si2} & 58  & -0.11 & 8.2 & - & -0.2 &  700\cite{CeCu2Si2ref}\\
  \cerhin\ (In(1)$_c$) & 12 & -2.51 & 26.0 & - & 1.3 &  200\cite{CeRhIn5ref}\\
  \cerhin\ (In(1)$_{ab}$) & 10  & -0.54 & 19.6 & - & 2.2 &  200\cite{CeRhIn5ref}\\
  CeAl$_3$ (Al)\cite{lysakCeAl3} & 60 & 0.02 & 3.5 & - & -0.7 &  1620\cite{CeAl3ref}\\
  CePtSi (Si)\cite{benliCePtSi} & 20 & -0.11 & 7.1 & - & -1.7 &  800\cite{Lee1987}\\
  CePtSi$_{0.9}$Ge$_{0.1}$ (Si)\cite{benliCePtSi} & 15 & 0.07 & 4.2 & - & -1.4 &  1350\cite{Steglich1994}\\
  CeSn$_3$ (Sn)\cite{malikCeSn3} & 167 & -0.05 & 32 & - &  0.2  &  70\cite{Stassis1981}\\
  Ce$_3$Bi$_4$Pt$_3$ (Bi)\cite{reyes343} & 123 & 0.37 & 46 & - & -1.0  &  3.3\cite{Hundley1990}\\
  YbCuAl (Cu)\cite{MacLaughlinYbCuAl} & 73 & 0.07 & -1.0 & - & 0.03 &  260\cite{Mattens1977}\\
  URu$_2$Si$_2$ (Si$_c$)\cite{bernal} & 84 & 0.05 & 3.37 & - & -0.03  &  65\cite{URu2Si2ref,URu2Si2ref2}\\
  CeP (P)\cite{narath} & 76 & 0.03 & 9.98 & - & -1.49  &  17\cite{Kwon1991}\\
  CeAs (As)\cite{narath} & 73 & 0.43 & 16.3 & - & -2.41  &  ? \\
  UPt$_3$ (Pt$_c$)\cite{halparin} & 23 & 3.95 & -95.7 & - & 0.19  &  420\cite{UPt3ref}\\
  UPt$_3$ (Pt$_{ab}$)\cite{halparin} & 19 & -2.0 & -54.4 & - & 1.30  &  420\cite{UPt3ref}\\
  UBe$_{13}$ (Be)\cite{ClarkUBe13} & 10 & -0.02 & 0.86 & - & -0.008  &  900\cite{UBe13ref}\\
%  \hline
\end{tabular}
\end{ruledtabular}
\end{table*}

Based on our observation of universality of $\chi _{\mathrm{cf}}$ we can now
make contact to the two fluid picture of NPF and demonstrate that indeed $%
\chi _{\mathrm{cf}}$ agrees with the predictions of their phenomenological
approach. In Ce$_{1-x}$La$_{x}$CoIn$_{5}$, NPF proposed that $\chi _{\mathrm{%
cf}}(T)=f(T)RC_{\mathrm{cf}}/T$, where $R$ is the Wilson ratio, which
successfully explains the doping evolution of the bulk properties.
Empirically it was found that $f(T)\sim 1-T/T^{\ast }$, and that $C_{\mathrm{%
cf}}/T\sim \log (T)$. Combining these results, we arrive at Eq. (\ref%
{eqn:chiHE}), a candidate description of $\chi _{\mathrm{cf}}(T)$. Indeed,
in the inserts of Figs. \ref{fig:In1CeCoIn5} and \ref{fig:In2CeCoIn5} we
show fits of $K_{\mathrm{cf}}(T)$ to the equation:
\begin{equation}
K_{\mathrm{cf}}(T)=K_{\mathrm{cf}}^{0}\left( 1-\frac{T}{T^{\ast }}\right)
\log \frac{T^{\ast }}{T},  \label{eqn:KKL}
\end{equation}%
with $K_{\mathrm{cf}}^{0}$ and $T^{\ast }$ as fitting parameters. $K_{%
\mathrm{cf}}^{0}$ is given by the value of the shift at $T=\alpha T^{\ast }$%
, where $\alpha \approx 0.259$ is given by the equation $(\alpha -1)\log
\alpha =1$. Fig. \ref{fig:KKLvsChiKL} shows $K_{\mathrm{cf}}$ versus $\chi _{%
\mathrm{cf}}$ in CeCoIn$_{5}$, where the $\chi _{\mathrm{cf}}$ data were
extracted from bulk measurements by NPF. \cite{NPF,satoru} The linearity,
especially for the $c$ direction, strongly supports the argument that the
second component measured by NMR Knight shifts is indeed probing $\chi _{%
\mathrm{cf}}$. Figs. \ref{fig:In1CeRhIn5} - \ref{fig:UPt3} present
comparable data for a number of Ce, Yb and U compounds. In Fig. \ref%
{fig:scalingplot} we show $K_{\mathrm{cf}}(T)/K_{\mathrm{cf}}^{0}$ versus $%
T/T^{\ast }$ for all of the materials for which we have thus far
been able to obtain Knight shift and susceptibility data. With the
exception of Ce$_{3} $Bi$_{4}$Pt$_{3}$, a Kondo insulator, and
CeSn$_{3}$, a mixed valent system, the data scale remarkably well
with one another. This result points to a common mechanism for the
Knight shift anomaly, a conclusion that is model independent.
 In fact, the scaling evident in Fig. \ref{fig:scalingplot} is
based solely on the reasonable assumption that a second component of
susceptibility with a different hyperfine coupling manifests itself below $%
T^{\ast }$, and that it is this component that exhibits the universal
scaling behavior given by Eq. (\ref{eqn:chiHE}). The fact that the scaling
form agrees with the analysis presented by NPF supports the
two fluid description for a broad range of heavy fermion
materials.

\subsection{Discussion and Conclusions}

The alert reader will have noticed that there are two materials,
Ce$_{3}$Bi$_{4}$Pt$_{3}$ and CeSn$_{3}$,  for which $K$ once more
becomes proportional to $\chi$ for $T<T_0$. This result has a
simple physical interpretation: $T_0$ marks the temperature below
which there are no longer any local moments present in the
material, so below $T_0$ the system reverts to a single component.
In Ce$_{3}$Bi$_{4}$Pt$_{3}$ that component becomes a Kondo
insulator, with a band gap in the one component electronic system
brought about by band structure effects. In the case of CeSn$_{3}$
that single component is a heavy fermion liquid. This point of
view finds support in the measurements of the specific heat and
resistivity of CeSn$_{3}$ which show Fermi liquid behavior below
$T\approx 17$K. A reasonable explanation is that for these
materials $f(T)$ reaches unity at $T_{0}$. Below this temperature,
the heavy electron liquid is fully formed, and there is only a
single, itinerant magnetic component.

In the other heavy fermion systems, both the localized and heavy
electron liquid components coexist down to the lowest temperatures
measured, typically defined by the onset of magnetic or
superconducting order in the particular compounds. Presumably, in
the absence of order and for sufficiently low temperatures all
materials should exhibit one component behavior corresponding to
the fact that the heavy electron emergence has become complete,
there is no further trace of local moment behavior, and $f=1$. In
fact, for CeCu$_{2}$Si$_{2}$, Fig. \ref{fig:scalingplot}
suggests that perhaps $K_{\mathrm{cf}}$ begins to saturates below $%
0.02T^{\ast }\cong 3.5$K.

The distinct $T$-dependence of \ $\chi _{\mathrm{cf%
}}$ and $\chi _{\mathrm{ff}}$ below $T^{\ast }$ \ is another strong
indication for the fact that $T^{\ast }$ is not the single ion Kondo
temperature $T_{K}$. In a few cases we were able to determine the actual
temperature dependence of $\chi _{\mathrm{ff}}$ below $T^{\ast }$. This is
possible if the low temperature bulk susceptibility clearly shows a
logarithmic temperature dependence allowing us to determine the unknown
hyperfine constant $A$. This approach \ yields a Curie-Weiss type
susceptibility for $\left\langle S_{f}S_{f}\right\rangle $ with Weiss
temperature equal to $T^{\ast }$ as determined in the fits for $\left\langle
S_{f}S_{c}\right\rangle $. This \ is yet another reason for the collective,
rather than local origin of $T^{\ast }$. Finally this point of view is
supported by the fact that the latter scenario typically leads to the same $%
T $-dependence of $\chi _{\mathrm{\mathrm{ff}}}$ and $\chi
_{\mathrm{cf}}$ below some coherence temperature.\cite{MillisLee}
The notion of a two fluid description to Kondo lattice systems
has, in some way, been discussed in theories based on single ion
dynamics\cite{MillisLee,Auerbach,Miyake,Miyake2} . The
unconventional temperature dependence of $\chi _{\mathrm{cf}}$
clearly requires a new approach to Kondo lattice systems which
goes beyond those theories.

For the materials that possess structural symmetries lower than cubic, we
note that $T^*$ is anisotropic, in some cases by more than a factor of two.
In the two-fluid model of NPF, $T^*$ is a measure of the Ce-Ce intersite
coupling. For the bulk measurements presented by NPF, the measured $T^*$
describes an volume average coupling. However NMR results probe a local
susceptibility, and the anisotropic $T^*$'s measured by NMR reflects the
anisotropy of the local couplings between the 4f (5f) sites. This anisotropy
reflects that of the orbitals that enter the quantum chemistry calculation
of the nearest neighbor coupling; in fact the anisotropy might be maximum
for directions intermediate to the $c$ and $ab$ planar directions.

We emphasize that the scaling behavior exhibited strongly supports
the validity of the two-fluid description in a wide variety of
Kondo lattice systems, ranging from heavy electron systems to
mixed valent systems. This scaling seems to be independent of the
ground state: the materials represented here include magnetically
ordered, superconducting, as well as Kondo insulating materials.

The specific heat of the heavy electron fluid of the system was shown by NPF
to behave as
\begin{equation}
C_{\mathrm{cf}}\left( T\right) =Q\frac{T}{T^{\ast }}f\left( T\right) \log
\left( \frac{T^{\ast }}{T}\right)
\end{equation}%
where the dimensionless constant $Q$ determines the entropy contribution of
the heavy electron fluid at $T^{\ast }$, $S_{\mathrm{cf}}\left( T^{\ast
}\right) =\int_{0}^{T^{\ast }}\frac{C_{\mathrm{cf}}\left( T\right) }{T}dT=%
\frac{3}{4}Q$ . It is natural to assume that
$S_{\mathrm{cf}}\left( T^{\ast }\right) $ is a generic value,
independent of the details of the system.
Together with the fixed Wilson ratio, $R=\frac{C_{\mathrm{cf}}}{T\chi _{%
\mathrm{cf}}}$, of the heavy electron component of the two fluid system,
this \ gives
\begin{equation}
\chi _{\mathrm{cf}}=\frac{4}{3R}\frac{S_{\mathrm{cf}}\left( T^{\ast }\right)
}{T^{\ast }}f\left( T\right) \log \left( \frac{T^{\ast }}{T}\right)
\end{equation}%
demonstrating that not only the $T$-dependence but also the absolute value
of $\chi _{\mathrm{cf}}$ is determined by $T^{\ast }$. In our current
analysis, we do not know the hyperfine coupling constant $A$ for all
materials and consequently can not determine the prefactor in $\chi _{%
\mathrm{cf}}$. In addition, even if there exists a generic value of $S_{%
\mathrm{cf}}\left( T^{\ast }\right) $, this does not, however, imply a
universal value for the specific heat coefficient $\gamma =\left. \frac{%
C\left( T\right) }{T}\right\vert _{T\rightarrow 0}$. A $\gamma $-value can
only be defined if at some low temperature $T_{0}\ll T^{\ast }$ the
logarithmic growth of $\frac{C_{\mathrm{cf}}\left( T\right) }{T}$ stops.
Then $\gamma \simeq \frac{Q}{T^{\ast }}\log \left( \frac{T^{\ast }}{T_{0}}%
\right) $ is determined by both $T^{\ast }$ and $T_{0}$. Such behavior might
reflect, for example, the crossover from a quantum critical regime to a
heavy Fermi liquid regime if the system is close to a quantum critical
point. As seen in Fig. \ref{fig:TstarVSgamma}, no correlation between $%
\gamma $ and $T^{\ast }$ exists. From the NPF perspective, this result is
not surprising.

It is interesting to note that for the Cu sites that are nearest
to the Fe impurities in the dilute Kondo alloy Cu$_{1-x}$Fe$_x$
there is no Knight shift anomaly.\cite{alloul,ishii} This result
supports the argument that the
Knight shift anomaly observed in Kondo lattice systems is a \textit{%
correlated} Kondo effect, determined by the onset of the heavy electron
component with a different hyperfine component, rather than a local property
of a Kondo screened impurity.

In summary, both the regime of uncorrelated local moments at very
high $T$, as well as the heavy Fermi liquid at $T<<T_K$ should
exhibit $K \propto \chi$ (with different proportionality
constants). At high $T$ local moments dominate the magnetic
response and the conduction electrons are invisible by comparison,
in the other limit a single component Fermi liquid state has
emerged. However, in the important regime in between, which is so
crucial to understand how a heavy electron emerges, and which
might dominate all the way to $T=0$ at a quantum critical point,
the two component picture is essential. It is this regime where
the systems are characterized by the scaling behavior found in
this paper. The unexpected simplicity captured by the two
component model offers many new opportunities for the
reinterpretation of existing data and future experiments in a
simple manner.

\acknowledgements{ This work was performed at Los Alamos National
Laboratory under the auspices of the US Department of Energy, and
has been supported by the research network on correlated matter of
the Institute for Complex Adaptive Matter of the University of
California. The authors thank Ar. Abanov, E. Abrahams, F. Borsa,
P. Canfield, P. Coleman, D. L. Cox ,Z. Fisk, R. Heffner, F.
Hellman, S. Kos, J. Mydosh, S. Nakatsuji, C. P. Slichter, and G.
Sparn for stimulating discussions.}

%\bibliographystyle{apsrev}
%\bibliography{heavyfermionbib3}

\begin{thebibliography}{50}
\expandafter\ifx\csname
natexlab\endcsname\relax\def\natexlab#1{#1}\fi
\expandafter\ifx\csname bibnamefont\endcsname\relax
  \def\bibnamefont#1{#1}\fi
\expandafter\ifx\csname bibfnamefont\endcsname\relax
  \def\bibfnamefont#1{#1}\fi
\expandafter\ifx\csname citenamefont\endcsname\relax
  \def\citenamefont#1{#1}\fi
\expandafter\ifx\csname url\endcsname\relax
  \def\url#1{\texttt{#1}}\fi
\expandafter\ifx\csname
urlprefix\endcsname\relax\def\urlprefix{URL }\fi
\providecommand{\bibinfo}[2]{#2}
\providecommand{\eprint}[2][]{\url{#2}}

\bibitem[{\citenamefont{Custers et~al.}(2003)\citenamefont{Custers, Gegenwart,
  Wilhelm, Neumaier, Tokiwa, Trovarelli, Geibel, Steglich, Pepin, and
  Coleman}}]{Custers}
\bibinfo{author}{\bibfnamefont{J.}~\bibnamefont{Custers}},
  \bibinfo{author}{\bibfnamefont{P.}~\bibnamefont{Gegenwart}},
  \bibinfo{author}{\bibfnamefont{H.}~\bibnamefont{Wilhelm}},
  \bibinfo{author}{\bibfnamefont{K.}~\bibnamefont{Neumaier}},
  \bibinfo{author}{\bibfnamefont{Y.}~\bibnamefont{Tokiwa}},
  \bibinfo{author}{\bibfnamefont{O.}~\bibnamefont{Trovarelli}},
  \bibinfo{author}{\bibfnamefont{C.}~\bibnamefont{Geibel}},
  \bibinfo{author}{\bibfnamefont{F.}~\bibnamefont{Steglich}},
  \bibinfo{author}{\bibfnamefont{C.}~\bibnamefont{Pepin}}, \bibnamefont{and}
  \bibinfo{author}{\bibfnamefont{P.}~\bibnamefont{Coleman}},
  \bibinfo{journal}{Nature}
  \textbf{\bibinfo{volume}{424}}(\bibinfo{number}{6948}), \bibinfo{pages}{524 }
  (\bibinfo{year}{2003}).

\bibitem[{\citenamefont{Kuchler et~al.}(2003)\citenamefont{Kuchler, Oeschler,
  Gegenwart, Cichorek, Neumaier, Tegus, Geibel, Mydosh, Steglich, Zhu
  et~al.}}]{Gegenward}
\bibinfo{author}{\bibfnamefont{R.}~\bibnamefont{Kuchler}},
  \bibinfo{author}{\bibfnamefont{N.}~\bibnamefont{Oeschler}},
  \bibinfo{author}{\bibfnamefont{P.}~\bibnamefont{Gegenwart}},
  \bibinfo{author}{\bibfnamefont{T.}~\bibnamefont{Cichorek}},
  \bibinfo{author}{\bibfnamefont{K.}~\bibnamefont{Neumaier}},
  \bibinfo{author}{\bibfnamefont{O.}~\bibnamefont{Tegus}},
  \bibinfo{author}{\bibfnamefont{C.}~\bibnamefont{Geibel}},
  \bibinfo{author}{\bibfnamefont{J.~A.} \bibnamefont{Mydosh}},
  \bibinfo{author}{\bibfnamefont{F.}~\bibnamefont{Steglich}},
  \bibinfo{author}{\bibfnamefont{L.}~\bibnamefont{Zhu}}, \bibnamefont{et~al.},
  \bibinfo{journal}{Phys. Rev. Lett.}
  \textbf{\bibinfo{volume}{91}}(\bibinfo{number}{6}), \bibinfo{pages}{066405 }
  (\bibinfo{year}{2003}).

\bibitem[{\citenamefont{Stewart}(1984)}]{Steward}
\bibinfo{author}{\bibfnamefont{G.~R.} \bibnamefont{Stewart}},
  \bibinfo{journal}{Rev. Mod. Phys.}
  \textbf{\bibinfo{volume}{56}}(\bibinfo{number}{4}), \bibinfo{pages}{755 }
  (\bibinfo{year}{1984}).

\bibitem[{\citenamefont{vonLohneysen}(1996)}]{HvL}
\bibinfo{author}{\bibfnamefont{H.}~\bibnamefont{vonLohneysen}},
  \bibinfo{journal}{Journal of Physics: Condensed Matter}
  \textbf{\bibinfo{volume}{8}}(\bibinfo{number}{48}), \bibinfo{pages}{9689 }
  (\bibinfo{year}{1996}).

\bibitem[{\citenamefont{Schroder et~al.}(2000)\citenamefont{Schroder, Aeppli,
  Coldea, Adams, Stockert, von Lohneysen, Bucher, Ramazashvili, and
  Coleman}}]{Schroder}
\bibinfo{author}{\bibfnamefont{A.}~\bibnamefont{Schroder}},
  \bibinfo{author}{\bibfnamefont{G.}~\bibnamefont{Aeppli}},
  \bibinfo{author}{\bibfnamefont{R.}~\bibnamefont{Coldea}},
  \bibinfo{author}{\bibfnamefont{M.}~\bibnamefont{Adams}},
  \bibinfo{author}{\bibfnamefont{O.}~\bibnamefont{Stockert}},
  \bibinfo{author}{\bibfnamefont{H.}~\bibnamefont{von Lohneysen}},
  \bibinfo{author}{\bibfnamefont{E.}~\bibnamefont{Bucher}},
  \bibinfo{author}{\bibfnamefont{R.}~\bibnamefont{Ramazashvili}},
  \bibnamefont{and} \bibinfo{author}{\bibfnamefont{P.}~\bibnamefont{Coleman}},
  \bibinfo{journal}{Nature}
  \textbf{\bibinfo{volume}{407}}(\bibinfo{number}{6802}), \bibinfo{pages}{351 }
  (\bibinfo{year}{2000}).

\bibitem[{\citenamefont{Si et~al.}(2001)\citenamefont{Si, Rabello, Ingersent,
  and Smith}}]{Si}
\bibinfo{author}{\bibfnamefont{Q.~M.} \bibnamefont{Si}},
  \bibinfo{author}{\bibfnamefont{S.}~\bibnamefont{Rabello}},
  \bibinfo{author}{\bibfnamefont{K.}~\bibnamefont{Ingersent}},
  \bibnamefont{and} \bibinfo{author}{\bibfnamefont{J.~L.} \bibnamefont{Smith}},
  \bibinfo{journal}{Nature}
  \textbf{\bibinfo{volume}{413}}(\bibinfo{number}{6858}), \bibinfo{pages}{804 }
  (\bibinfo{year}{2001}).

\bibitem[{\citenamefont{Coleman
  et~al.}(2000{\natexlab{a}})\citenamefont{Coleman, Pepin, and
  Tsvelik}}]{Piers}
\bibinfo{author}{\bibfnamefont{P.}~\bibnamefont{Coleman}},
  \bibinfo{author}{\bibfnamefont{C.}~\bibnamefont{Pepin}}, \bibnamefont{and}
  \bibinfo{author}{\bibfnamefont{A.~M.} \bibnamefont{Tsvelik}},
  \bibinfo{journal}{Phys. Rev. B}
  \textbf{\bibinfo{volume}{62}}(\bibinfo{number}{6}), \bibinfo{pages}{3852 }
  (\bibinfo{year}{2000}{\natexlab{a}}).

\bibitem[{\citenamefont{Coleman
  et~al.}(2000{\natexlab{b}})\citenamefont{Coleman, Pepin, and
  Tsvelik}}]{Piers2}
\bibinfo{author}{\bibfnamefont{P.}~\bibnamefont{Coleman}},
  \bibinfo{author}{\bibfnamefont{C.}~\bibnamefont{Pepin}}, \bibnamefont{and}
  \bibinfo{author}{\bibfnamefont{A.~M.} \bibnamefont{Tsvelik}},
  \bibinfo{journal}{Nuclear Phys. B}
  \textbf{\bibinfo{volume}{586}}(\bibinfo{number}{3}), \bibinfo{pages}{641 }
  (\bibinfo{year}{2000}{\natexlab{b}}).

\bibitem[{\citenamefont{Jerez et~al.}(2003)\citenamefont{Jerez, Lavagna, and
  Bensimon}}]{Lavagna}
\bibinfo{author}{\bibfnamefont{A.}~\bibnamefont{Jerez}},
  \bibinfo{author}{\bibfnamefont{M.}~\bibnamefont{Lavagna}}, \bibnamefont{and}
  \bibinfo{author}{\bibfnamefont{D.}~\bibnamefont{Bensimon}},
  \bibinfo{journal}{Phys. Rev. B}
  \textbf{\bibinfo{volume}{68}}(\bibinfo{number}{9}), \bibinfo{pages}{94410 }
  (\bibinfo{year}{2003}).

\bibitem[{\citenamefont{Senthil et~al.}(2004)\citenamefont{Senthil, Vojta, and
  Sachdev}}]{Sachdev}
\bibinfo{author}{\bibfnamefont{T.}~\bibnamefont{Senthil}},
  \bibinfo{author}{\bibfnamefont{M.}~\bibnamefont{Vojta}}, \bibnamefont{and}
  \bibinfo{author}{\bibfnamefont{S.}~\bibnamefont{Sachdev}},
  \bibinfo{journal}{Phys. Rev. B}
  \textbf{\bibinfo{volume}{69}}(\bibinfo{number}{3}), \bibinfo{pages}{035111}
  (\bibinfo{year}{2004}).

\bibitem[{\citenamefont{Myers and Narath}(1973)}]{narath}
\bibinfo{author}{\bibfnamefont{S.}~\bibnamefont{Myers}} \bibnamefont{and}
  \bibinfo{author}{\bibfnamefont{A.}~\bibnamefont{Narath}},
  \bibinfo{journal}{Solid State Communications}
  \textbf{\bibinfo{volume}{12}}(\bibinfo{number}{1}), \bibinfo{pages}{83 }
  (\bibinfo{year}{1973}).

\bibitem[{\citenamefont{MacLaughlin}(1985)}]{maclaughlinreview}
\bibinfo{author}{\bibfnamefont{D.}~\bibnamefont{MacLaughlin}},
  \bibinfo{journal}{Journal of Magnetism and Magnetic Materials}
  \textbf{\bibinfo{volume}{47-48}}, \bibinfo{pages}{121 }
  (\bibinfo{year}{1985}).

\bibitem[{\citenamefont{Ohama et~al.}(1995)\citenamefont{Ohama, Yasuoka,
  Mandrus, Fisk, and Smith}}]{yasuokaCeCu2Si2}
\bibinfo{author}{\bibfnamefont{T.}~\bibnamefont{Ohama}},
  \bibinfo{author}{\bibfnamefont{H.}~\bibnamefont{Yasuoka}},
  \bibinfo{author}{\bibfnamefont{D.}~\bibnamefont{Mandrus}},
  \bibinfo{author}{\bibfnamefont{Z.}~\bibnamefont{Fisk}}, \bibnamefont{and}
  \bibinfo{author}{\bibfnamefont{J.~L.} \bibnamefont{Smith}},
  \bibinfo{journal}{J. Phys. Soc. Jpn.} \textbf{\bibinfo{volume}{64}},
  \bibinfo{pages}{2628 } (\bibinfo{year}{1995}).

\bibitem[{\citenamefont{Kim et~al.}(1995)\citenamefont{Kim, Makivic, and
  Cox}}]{cox}
\bibinfo{author}{\bibfnamefont{E.}~\bibnamefont{Kim}},
  \bibinfo{author}{\bibfnamefont{M.}~\bibnamefont{Makivic}}, \bibnamefont{and}
  \bibinfo{author}{\bibfnamefont{D.~L.} \bibnamefont{Cox}},
  \bibinfo{journal}{Phys. Rev. Lett.} \textbf{\bibinfo{volume}{75}},
  \bibinfo{pages}{2015 } (\bibinfo{year}{1995}).

\bibitem[{\citenamefont{Sonier et~al.}(2000)\citenamefont{Sonier, Heffner,
  MacLaughlin, Smith, Cooley, and Nieuwenhuys}}]{sonier}
\bibinfo{author}{\bibfnamefont{J.}~\bibnamefont{Sonier}},
  \bibinfo{author}{\bibfnamefont{R.}~\bibnamefont{Heffner}},
  \bibinfo{author}{\bibfnamefont{D.}~\bibnamefont{MacLaughlin}},
  \bibinfo{author}{\bibfnamefont{J.}~\bibnamefont{Smith}},
  \bibinfo{author}{\bibfnamefont{J.}~\bibnamefont{Cooley}}, \bibnamefont{and}
  \bibinfo{author}{\bibfnamefont{G.}~\bibnamefont{Nieuwenhuys}},
  \bibinfo{journal}{Physica B} \textbf{\bibinfo{volume}{289-290}},
  \bibinfo{pages}{20 } (\bibinfo{year}{2000}).

\bibitem[{\citenamefont{Nakatsuji et~al.}(2004)\citenamefont{Nakatsuji, Pines,
  and Fisk}}]{NPF}
\bibinfo{author}{\bibfnamefont{S.}~\bibnamefont{Nakatsuji}},
  \bibinfo{author}{\bibfnamefont{D.}~\bibnamefont{Pines}}, \bibnamefont{and}
  \bibinfo{author}{\bibfnamefont{Z.}~\bibnamefont{Fisk}},
  \bibinfo{journal}{Physical Review Letters}
  \textbf{\bibinfo{volume}{92}}(\bibinfo{number}{1}), \bibinfo{pages}{016401/1
  } (\bibinfo{year}{2004}).

\bibitem[{\citenamefont{Curro et~al.}(2001)\citenamefont{Curro, Simovic,
  Hammel, Pagliuso, Sarrao, Thompson, and Martins}}]{curroCeCoIn5}
\bibinfo{author}{\bibfnamefont{N.~J.} \bibnamefont{Curro}},
  \bibinfo{author}{\bibfnamefont{B.}~\bibnamefont{Simovic}},
  \bibinfo{author}{\bibfnamefont{P.~C.} \bibnamefont{Hammel}},
  \bibinfo{author}{\bibfnamefont{P.~G.} \bibnamefont{Pagliuso}},
  \bibinfo{author}{\bibfnamefont{J.~L.} \bibnamefont{Sarrao}},
  \bibinfo{author}{\bibfnamefont{J.~D.} \bibnamefont{Thompson}},
  \bibnamefont{and} \bibinfo{author}{\bibfnamefont{G.~B.}
  \bibnamefont{Martins}}, \bibinfo{journal}{Phys. Rev. B}
  \textbf{\bibinfo{volume}{64}}, \bibinfo{pages}{180514}
  (\bibinfo{year}{2001}).

\bibitem[{\citenamefont{Mila and Rice}(1989)}]{milarice}
\bibinfo{author}{\bibfnamefont{F.}~\bibnamefont{Mila}} \bibnamefont{and}
  \bibinfo{author}{\bibfnamefont{T.}~\bibnamefont{Rice}},
  \bibinfo{journal}{Physica C}
  \textbf{\bibinfo{volume}{157}}(\bibinfo{number}{3}), \bibinfo{pages}{561 }
  (\bibinfo{year}{1989}).

\bibitem[{\citenamefont{Barzykin and Affleck}(1996)}]{BarzykinAffleck}
\bibinfo{author}{\bibfnamefont{V.}~\bibnamefont{Barzykin}} \bibnamefont{and}
  \bibinfo{author}{\bibfnamefont{I.}~\bibnamefont{Affleck}},
  \bibinfo{journal}{Phys. Rev. Lett.} \textbf{\bibinfo{volume}{76}},
  \bibinfo{pages}{4959} (\bibinfo{year}{1996}).

\bibitem[{\citenamefont{Petrovic et~al.}(2001)\citenamefont{Petrovic, Pagliuso,
  Hundley, Movshovich, Sarrao, Thompson, Fisk, and Monthoux}}]{CeCoIn5ref}
\bibinfo{author}{\bibfnamefont{C.}~\bibnamefont{Petrovic}},
  \bibinfo{author}{\bibfnamefont{P.}~\bibnamefont{Pagliuso}},
  \bibinfo{author}{\bibfnamefont{M.}~\bibnamefont{Hundley}},
  \bibinfo{author}{\bibfnamefont{R.}~\bibnamefont{Movshovich}},
  \bibinfo{author}{\bibfnamefont{J.}~\bibnamefont{Sarrao}},
  \bibinfo{author}{\bibfnamefont{J.}~\bibnamefont{Thompson}},
  \bibinfo{author}{\bibfnamefont{Z.}~\bibnamefont{Fisk}}, \bibnamefont{and}
  \bibinfo{author}{\bibfnamefont{P.}~\bibnamefont{Monthoux}},
  \bibinfo{journal}{Journal of Physics: Condensed Matter}
  \textbf{\bibinfo{volume}{13}}(\bibinfo{number}{17}), \bibinfo{pages}{L337 }
  (\bibinfo{year}{2001}).

\bibitem[{\citenamefont{Steglich et~al.}(1979)\citenamefont{Steglich, Aarts,
  Bredl, Lieke, Meschede, Franz, and Sch\"{a}fer}}]{CeCu2Si2ref}
\bibinfo{author}{\bibfnamefont{F.}~\bibnamefont{Steglich}},
  \bibinfo{author}{\bibfnamefont{J.}~\bibnamefont{Aarts}},
  \bibinfo{author}{\bibfnamefont{C.~D.} \bibnamefont{Bredl}},
  \bibinfo{author}{\bibfnamefont{W.}~\bibnamefont{Lieke}},
  \bibinfo{author}{\bibfnamefont{D.}~\bibnamefont{Meschede}},
  \bibinfo{author}{\bibfnamefont{W.}~\bibnamefont{Franz}}, \bibnamefont{and}
  \bibinfo{author}{\bibfnamefont{H.}~\bibnamefont{Sch\"{a}fer}},
  \bibinfo{journal}{Phys. Rev. Lett.} \textbf{\bibinfo{volume}{43}},
  \bibinfo{pages}{1892} (\bibinfo{year}{1979}).

\bibitem[{\citenamefont{Fisher et~al.}(2002)\citenamefont{Fisher, Bouquet,
  Phillips, Hundley, Pagliuso, Sarrao, Fisk, and Thompson}}]{CeRhIn5ref}
\bibinfo{author}{\bibfnamefont{A.}~\bibnamefont{Fisher}},
  \bibinfo{author}{\bibfnamefont{F.}~\bibnamefont{Bouquet}},
  \bibinfo{author}{\bibfnamefont{N.~E.} \bibnamefont{Phillips}},
  \bibinfo{author}{\bibfnamefont{M.~F.} \bibnamefont{Hundley}},
  \bibinfo{author}{\bibfnamefont{P.~G.} \bibnamefont{Pagliuso}},
  \bibinfo{author}{\bibfnamefont{J.~L.} \bibnamefont{Sarrao}},
  \bibinfo{author}{\bibfnamefont{Z.}~\bibnamefont{Fisk}}, \bibnamefont{and}
  \bibinfo{author}{\bibfnamefont{J.~D.} \bibnamefont{Thompson}},
  \bibinfo{journal}{Phys. Rev. B.} \textbf{\bibinfo{volume}{65}},
  \bibinfo{pages}{224509} (\bibinfo{year}{2002}).

\bibitem[{\citenamefont{Lysak and MacLaughlin}(1985)}]{lysakCeAl3}
\bibinfo{author}{\bibfnamefont{M.}~\bibnamefont{Lysak}} \bibnamefont{and}
  \bibinfo{author}{\bibfnamefont{D.}~\bibnamefont{MacLaughlin}},
  \bibinfo{journal}{Physical Review B (Condensed Matter)}
  \textbf{\bibinfo{volume}{31}}(\bibinfo{number}{11}), \bibinfo{pages}{6963 }
  (\bibinfo{year}{1985}).

\bibitem[{\citenamefont{Andres et~al.}(1975)\citenamefont{Andres, Graebner, and
  Ott}}]{CeAl3ref}
\bibinfo{author}{\bibfnamefont{K.}~\bibnamefont{Andres}},
  \bibinfo{author}{\bibfnamefont{J.}~\bibnamefont{Graebner}}, \bibnamefont{and}
  \bibinfo{author}{\bibfnamefont{H.}~\bibnamefont{Ott}},
  \bibinfo{journal}{Phys. Rev. Lett.}
  \textbf{\bibinfo{volume}{35}}(\bibinfo{number}{26}), \bibinfo{pages}{1779 }
  (\bibinfo{year}{1975}).

\bibitem[{\citenamefont{Young}(2003)}]{benliCePtSi}
\bibinfo{author}{\bibfnamefont{B.-L.} \bibnamefont{Young}}, Ph.D. thesis,
  \bibinfo{school}{Univ. of Calif., Riverside} (\bibinfo{year}{2003}).

\bibitem[{\citenamefont{Lee and Shelton}(1987)}]{Lee1987}
\bibinfo{author}{\bibfnamefont{W.}~\bibnamefont{Lee}} \bibnamefont{and}
  \bibinfo{author}{\bibfnamefont{R.}~\bibnamefont{Shelton}},
  \bibinfo{journal}{Physical Review B (Condensed Matter)}
  \textbf{\bibinfo{volume}{35}}(\bibinfo{number}{10}), \bibinfo{pages}{5369 }
  (\bibinfo{year}{1987}).

\bibitem[{\citenamefont{Steglich et~al.}(1994)\citenamefont{Steglich, Geibel,
  Gloos, Olesch, Schank, Wassilew, Loidl, Krimmel, and Stewart}}]{Steglich1994}
\bibinfo{author}{\bibfnamefont{F.}~\bibnamefont{Steglich}},
  \bibinfo{author}{\bibfnamefont{C.}~\bibnamefont{Geibel}},
  \bibinfo{author}{\bibfnamefont{K.}~\bibnamefont{Gloos}},
  \bibinfo{author}{\bibfnamefont{G.}~\bibnamefont{Olesch}},
  \bibinfo{author}{\bibfnamefont{C.}~\bibnamefont{Schank}},
  \bibinfo{author}{\bibfnamefont{C.}~\bibnamefont{Wassilew}},
  \bibinfo{author}{\bibfnamefont{A.}~\bibnamefont{Loidl}},
  \bibinfo{author}{\bibfnamefont{A.}~\bibnamefont{Krimmel}}, \bibnamefont{and}
  \bibinfo{author}{\bibfnamefont{G.}~\bibnamefont{Stewart}},
  \bibinfo{journal}{Journal of Low Temperature Physics}
  \textbf{\bibinfo{volume}{95}}(\bibinfo{number}{1-2}), \bibinfo{pages}{3 }
  (\bibinfo{year}{1994}).

\bibitem[{\citenamefont{Malik et~al.}(1975)\citenamefont{Malik, Vijayaraghavan,
  Garg, and Ripmeester}}]{malikCeSn3}
\bibinfo{author}{\bibfnamefont{S.~K.} \bibnamefont{Malik}},
  \bibinfo{author}{\bibfnamefont{R.}~\bibnamefont{Vijayaraghavan}},
  \bibinfo{author}{\bibfnamefont{S.~K.} \bibnamefont{Garg}}, \bibnamefont{and}
  \bibinfo{author}{\bibfnamefont{R.~J.} \bibnamefont{Ripmeester}},
  \bibinfo{journal}{Physica Status Solidi B} \textbf{\bibinfo{volume}{68}},
  \bibinfo{pages}{399 } (\bibinfo{year}{1975}).

\bibitem[{\citenamefont{Stassis et~al.}(1981)\citenamefont{Stassis, Loong,
  Zarestky, McMasters, and Nicklow}}]{Stassis1981}
\bibinfo{author}{\bibfnamefont{C.}~\bibnamefont{Stassis}},
  \bibinfo{author}{\bibfnamefont{C.-K.} \bibnamefont{Loong}},
  \bibinfo{author}{\bibfnamefont{J.}~\bibnamefont{Zarestky}},
  \bibinfo{author}{\bibfnamefont{O.}~\bibnamefont{McMasters}},
  \bibnamefont{and} \bibinfo{author}{\bibfnamefont{R.}~\bibnamefont{Nicklow}},
  \bibinfo{journal}{Physical Review B (Condensed Matter)}
  \textbf{\bibinfo{volume}{23}}(\bibinfo{number}{10}), \bibinfo{pages}{5128 }
  (\bibinfo{year}{1981}).

\bibitem[{\citenamefont{Reyes et~al.}(1994)\citenamefont{Reyes, Heffner,
  Canfield, Thompson, and Fisk}}]{reyes343}
\bibinfo{author}{\bibfnamefont{A.}~\bibnamefont{Reyes}},
  \bibinfo{author}{\bibfnamefont{R.}~\bibnamefont{Heffner}},
  \bibinfo{author}{\bibfnamefont{P.}~\bibnamefont{Canfield}},
  \bibinfo{author}{\bibfnamefont{J.}~\bibnamefont{Thompson}}, \bibnamefont{and}
  \bibinfo{author}{\bibfnamefont{Z.}~\bibnamefont{Fisk}},
  \bibinfo{journal}{Physical Review B (Condensed Matter)}
  \textbf{\bibinfo{volume}{49}}(\bibinfo{number}{23}), \bibinfo{pages}{16321 }
  (\bibinfo{year}{1994}).

\bibitem[{\citenamefont{Hundley et~al.}(1990)\citenamefont{Hundley, Canfield,
  Thompson, Fisk, and Lawrence}}]{Hundley1990}
\bibinfo{author}{\bibfnamefont{M.}~\bibnamefont{Hundley}},
  \bibinfo{author}{\bibfnamefont{P.}~\bibnamefont{Canfield}},
  \bibinfo{author}{\bibfnamefont{J.}~\bibnamefont{Thompson}},
  \bibinfo{author}{\bibfnamefont{Z.}~\bibnamefont{Fisk}}, \bibnamefont{and}
  \bibinfo{author}{\bibfnamefont{J.}~\bibnamefont{Lawrence}},
  \bibinfo{journal}{Physical Review B (Condensed Matter)}
  \textbf{\bibinfo{volume}{42}}(\bibinfo{number}{10}), \bibinfo{pages}{6842 }
  (\bibinfo{year}{1990}).

\bibitem[{\citenamefont{MacLaughlin et~al.}(1979)\citenamefont{MacLaughlin,
  de~Boer, Bijvoet, de~Chatel, and Mattens}}]{MacLaughlinYbCuAl}
\bibinfo{author}{\bibfnamefont{D.}~\bibnamefont{MacLaughlin}},
  \bibinfo{author}{\bibfnamefont{F.}~\bibnamefont{de~Boer}},
  \bibinfo{author}{\bibfnamefont{J.}~\bibnamefont{Bijvoet}},
  \bibinfo{author}{\bibfnamefont{P.}~\bibnamefont{de~Chatel}},
  \bibnamefont{and} \bibinfo{author}{\bibfnamefont{W.}~\bibnamefont{Mattens}},
  \bibinfo{journal}{Journal of Applied Physics}
  \textbf{\bibinfo{volume}{50}}(\bibinfo{number}{3 pt.2}), \bibinfo{pages}{2094
  } (\bibinfo{year}{1979}).

\bibitem[{\citenamefont{Mattens et~al.}(1977)\citenamefont{Mattens, Elenbaas,
  and de~Boer}}]{Mattens1977}
\bibinfo{author}{\bibfnamefont{W.}~\bibnamefont{Mattens}},
  \bibinfo{author}{\bibfnamefont{R.}~\bibnamefont{Elenbaas}}, \bibnamefont{and}
  \bibinfo{author}{\bibfnamefont{F.}~\bibnamefont{de~Boer}},
  \bibinfo{journal}{Communications on Physics}
  \textbf{\bibinfo{volume}{2}}(\bibinfo{number}{5}), \bibinfo{pages}{147 }
  (\bibinfo{year}{1977}).

\bibitem[{\citenamefont{Bernal et~al.}(2000)\citenamefont{Bernal, Becker,
  Mydosh, Nieuwenhuys, Menovsky, Paulus, Brom, MacLaughlin, and
  Lukefahr}}]{bernal}
\bibinfo{author}{\bibfnamefont{O.}~\bibnamefont{Bernal}},
  \bibinfo{author}{\bibfnamefont{B.}~\bibnamefont{Becker}},
  \bibinfo{author}{\bibfnamefont{J.}~\bibnamefont{Mydosh}},
  \bibinfo{author}{\bibfnamefont{G.}~\bibnamefont{Nieuwenhuys}},
  \bibinfo{author}{\bibfnamefont{A.}~\bibnamefont{Menovsky}},
  \bibinfo{author}{\bibfnamefont{P.}~\bibnamefont{Paulus}},
  \bibinfo{author}{\bibfnamefont{H.}~\bibnamefont{Brom}},
  \bibinfo{author}{\bibfnamefont{D.}~\bibnamefont{MacLaughlin}},
  \bibnamefont{and} \bibinfo{author}{\bibfnamefont{H.}~\bibnamefont{Lukefahr}},
  \bibinfo{journal}{Physica B} \textbf{\bibinfo{volume}{281-282}},
  \bibinfo{pages}{236 } (\bibinfo{year}{2000}).

\bibitem[{\citenamefont{Palstra et~al.}(1985)\citenamefont{Palstra, Menovsky,
  van~den Berg, Dirkmaat, Kes, Nieuwenhuys, and Mydosh}}]{URu2Si2ref}
\bibinfo{author}{\bibfnamefont{T.}~\bibnamefont{Palstra}},
  \bibinfo{author}{\bibfnamefont{A.}~\bibnamefont{Menovsky}},
  \bibinfo{author}{\bibfnamefont{J.}~\bibnamefont{van~den Berg}},
  \bibinfo{author}{\bibfnamefont{A.}~\bibnamefont{Dirkmaat}},
  \bibinfo{author}{\bibfnamefont{P.}~\bibnamefont{Kes}},
  \bibinfo{author}{\bibfnamefont{G.}~\bibnamefont{Nieuwenhuys}},
  \bibnamefont{and} \bibinfo{author}{\bibfnamefont{J.}~\bibnamefont{Mydosh}},
  \bibinfo{journal}{Phys. Rev. Lett.}
  \textbf{\bibinfo{volume}{55}}(\bibinfo{number}{24}), \bibinfo{pages}{2727 }
  (\bibinfo{year}{1985}).

\bibitem[{\citenamefont{Maple et~al.}(1986)\citenamefont{Maple, Chen,
  Dalichaouch, Kohara, Rossel, Torikachvili, McElfresh, and
  Thompson}}]{URu2Si2ref2}
\bibinfo{author}{\bibfnamefont{M.}~\bibnamefont{Maple}},
  \bibinfo{author}{\bibfnamefont{J.}~\bibnamefont{Chen}},
  \bibinfo{author}{\bibfnamefont{Y.}~\bibnamefont{Dalichaouch}},
  \bibinfo{author}{\bibfnamefont{T.}~\bibnamefont{Kohara}},
  \bibinfo{author}{\bibfnamefont{C.}~\bibnamefont{Rossel}},
  \bibinfo{author}{\bibfnamefont{M.}~\bibnamefont{Torikachvili}},
  \bibinfo{author}{\bibfnamefont{M.}~\bibnamefont{McElfresh}},
  \bibnamefont{and} \bibinfo{author}{\bibfnamefont{J.}~\bibnamefont{Thompson}},
  \bibinfo{journal}{Phys. Rev. Lett.}
  \textbf{\bibinfo{volume}{56}}(\bibinfo{number}{2}), \bibinfo{pages}{185 }
  (\bibinfo{year}{1986}).

\bibitem[{\citenamefont{Kwon et~al.}(1991)\citenamefont{Kwon, Haga, Nakamura,
  Suzuki, and Kasuya}}]{Kwon1991}
\bibinfo{author}{\bibfnamefont{Y.}~\bibnamefont{Kwon}},
  \bibinfo{author}{\bibfnamefont{Y.}~\bibnamefont{Haga}},
  \bibinfo{author}{\bibfnamefont{O.}~\bibnamefont{Nakamura}},
  \bibinfo{author}{\bibfnamefont{T.}~\bibnamefont{Suzuki}}, \bibnamefont{and}
  \bibinfo{author}{\bibfnamefont{T.}~\bibnamefont{Kasuya}},
  \bibinfo{journal}{Physica B}
  \textbf{\bibinfo{volume}{171}}(\bibinfo{number}{1-4}), \bibinfo{pages}{324 }
  (\bibinfo{year}{1991}).

\bibitem[{\citenamefont{Lee et~al.}(1993)\citenamefont{Lee, Moores, Song,
  Halperin, Kim, and Stewart}}]{halparin}
\bibinfo{author}{\bibfnamefont{M.}~\bibnamefont{Lee}},
  \bibinfo{author}{\bibfnamefont{G.}~\bibnamefont{Moores}},
  \bibinfo{author}{\bibfnamefont{Y.-Q.} \bibnamefont{Song}},
  \bibinfo{author}{\bibfnamefont{W.}~\bibnamefont{Halperin}},
  \bibinfo{author}{\bibfnamefont{W.}~\bibnamefont{Kim}}, \bibnamefont{and}
  \bibinfo{author}{\bibfnamefont{G.}~\bibnamefont{Stewart}},
  \bibinfo{journal}{Physical Review B (Condensed Matter)}
  \textbf{\bibinfo{volume}{48}}(\bibinfo{number}{10}), \bibinfo{pages}{7392 }
  (\bibinfo{year}{1993}).

\bibitem[{\citenamefont{Stewart et~al.}(1984)\citenamefont{Stewart, Fisk,
  Willis, and Smith}}]{UPt3ref}
\bibinfo{author}{\bibfnamefont{G.~R.} \bibnamefont{Stewart}},
  \bibinfo{author}{\bibfnamefont{Z.}~\bibnamefont{Fisk}},
  \bibinfo{author}{\bibfnamefont{J.~O.} \bibnamefont{Willis}},
  \bibnamefont{and} \bibinfo{author}{\bibfnamefont{J.~L.} \bibnamefont{Smith}},
  \bibinfo{journal}{Phys. Rev. Lett.} \textbf{\bibinfo{volume}{52}},
  \bibinfo{pages}{679} (\bibinfo{year}{1984}).

\bibitem[{\citenamefont{Clark et~al.}(1987)\citenamefont{Clark, Lan, van
  Kalkeren, Wong, Tien, Maclaughlin, Smith, Fisk, and Ott}}]{ClarkUBe13}
\bibinfo{author}{\bibfnamefont{W.}~\bibnamefont{Clark}},
  \bibinfo{author}{\bibfnamefont{M.}~\bibnamefont{Lan}},
  \bibinfo{author}{\bibfnamefont{G.}~\bibnamefont{van Kalkeren}},
  \bibinfo{author}{\bibfnamefont{W.}~\bibnamefont{Wong}},
  \bibinfo{author}{\bibfnamefont{C.}~\bibnamefont{Tien}},
  \bibinfo{author}{\bibfnamefont{D.}~\bibnamefont{Maclaughlin}},
  \bibinfo{author}{\bibfnamefont{J.}~\bibnamefont{Smith}},
  \bibinfo{author}{\bibfnamefont{Z.}~\bibnamefont{Fisk}}, \bibnamefont{and}
  \bibinfo{author}{\bibfnamefont{H.}~\bibnamefont{Ott}},
  \bibinfo{journal}{Journal of Magnetism and Magnetic Materials}
  \textbf{\bibinfo{volume}{63-64}}, \bibinfo{pages}{396 }
  (\bibinfo{year}{1987}).

\bibitem[{\citenamefont{Ott et~al.}(1984)\citenamefont{Ott, Rudigier, Fisk, and
  Smith}}]{UBe13ref}
\bibinfo{author}{\bibfnamefont{H.~R.} \bibnamefont{Ott}},
  \bibinfo{author}{\bibfnamefont{H.}~\bibnamefont{Rudigier}},
  \bibinfo{author}{\bibfnamefont{Z.}~\bibnamefont{Fisk}}, \bibnamefont{and}
  \bibinfo{author}{\bibfnamefont{J.~L.} \bibnamefont{Smith}},
  \bibinfo{journal}{Physica} \textbf{\bibinfo{volume}{127B}},
  \bibinfo{pages}{359} (\bibinfo{year}{1984}).

\bibitem[{\citenamefont{Nakatsuji}(2003)}]{satoru}
\bibinfo{author}{\bibfnamefont{S.}~\bibnamefont{Nakatsuji}},
  \bibinfo{journal}{private communication}  (\bibinfo{year}{2003}).

\bibitem[{\citenamefont{Millis and Lee}(1987)}]{MillisLee}
\bibinfo{author}{\bibfnamefont{A.~J.} \bibnamefont{Millis}} \bibnamefont{and}
  \bibinfo{author}{\bibfnamefont{P.~A.} \bibnamefont{Lee}},
  \bibinfo{journal}{Phys. Rev. B} \textbf{\bibinfo{volume}{35}},
  \bibinfo{pages}{3394} (\bibinfo{year}{1987}).

\bibitem[{\citenamefont{Auerbach and Levin}(1986)}]{Auerbach}
\bibinfo{author}{\bibfnamefont{A.}~\bibnamefont{Auerbach}} \bibnamefont{and}
  \bibinfo{author}{\bibfnamefont{K.}~\bibnamefont{Levin}},
  \bibinfo{journal}{Phys. Rev. Lett.}
  \textbf{\bibinfo{volume}{57}}(\bibinfo{number}{7}), \bibinfo{pages}{877 }
  (\bibinfo{year}{1986}).

\bibitem[{\citenamefont{Kuramoto and Miyake}(1990)}]{Miyake}
\bibinfo{author}{\bibfnamefont{Y.}~\bibnamefont{Kuramoto}} \bibnamefont{and}
  \bibinfo{author}{\bibfnamefont{K.}~\bibnamefont{Miyake}},
  \bibinfo{journal}{J. Phys. Soc. Jpn.}
  \textbf{\bibinfo{volume}{59}}(\bibinfo{number}{8}), \bibinfo{pages}{2831 }
  (\bibinfo{year}{1990}).

\bibitem[{\citenamefont{Kuramoto and Miyake}(1992)}]{Miyake2}
\bibinfo{author}{\bibfnamefont{Y.}~\bibnamefont{Kuramoto}} \bibnamefont{and}
  \bibinfo{author}{\bibfnamefont{K.}~\bibnamefont{Miyake}},
  \bibinfo{journal}{Prog. Theor. Phys. Suppl.} (\bibinfo{number}{108}),
  \bibinfo{pages}{199 } (\bibinfo{year}{1992}).

\bibitem[{\citenamefont{Alloul}(1977)}]{alloul}
\bibinfo{author}{\bibfnamefont{H.}~\bibnamefont{Alloul}},
  \bibinfo{journal}{Physica} \textbf{\bibinfo{volume}{86-88B}},
  \bibinfo{pages}{449} (\bibinfo{year}{1977}).

\bibitem[{\citenamefont{Ishii}(1977)}]{ishii}
\bibinfo{author}{\bibfnamefont{H.}~\bibnamefont{Ishii}},
  \bibinfo{journal}{Physica} \textbf{\bibinfo{volume}{86-88B}},
  \bibinfo{pages}{517} (\bibinfo{year}{1977}).

\bibitem[{\citenamefont{Cooper et~al.}(1971)\citenamefont{Cooper, Rizzuto, and
  Olcese}}]{CooperCeSn3}
\bibinfo{author}{\bibfnamefont{J.~R.} \bibnamefont{Cooper}},
  \bibinfo{author}{\bibfnamefont{C.}~\bibnamefont{Rizzuto}}, \bibnamefont{and}
  \bibinfo{author}{\bibfnamefont{G.}~\bibnamefont{Olcese}},
  \bibinfo{journal}{Journal de Physique}
  \textbf{\bibinfo{volume}{32}}(\bibinfo{number}{2-3}), \bibinfo{pages}{C1 1136
  } (\bibinfo{year}{1971}).

\bibitem[{\citenamefont{Franse et~al.}(1985)\citenamefont{Franse, de~Visser,
  Menovsky, and Frings}}]{deVisser}
\bibinfo{author}{\bibfnamefont{J.}~\bibnamefont{Franse}},
  \bibinfo{author}{\bibfnamefont{A.}~\bibnamefont{de~Visser}},
  \bibinfo{author}{\bibfnamefont{A.}~\bibnamefont{Menovsky}}, \bibnamefont{and}
  \bibinfo{author}{\bibfnamefont{P.}~\bibnamefont{Frings}},
  \bibinfo{journal}{Journal of Magnetism and Magnetic Materials}
  \textbf{\bibinfo{volume}{52}}(\bibinfo{number}{1-4}), \bibinfo{pages}{61 }
  (\bibinfo{year}{1985}).

\end{thebibliography}

\newpage

\appendix

\subsection*{Appendix}

\begin{figure}[h]
\centering \includegraphics[width=\linewidth]{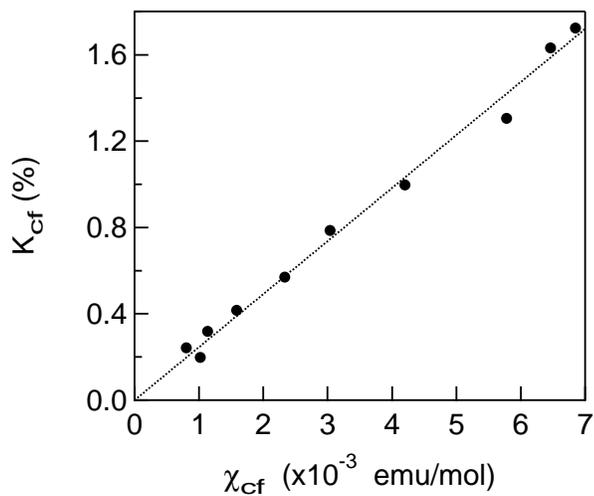}
\caption{$K_\mathrm{cf}$ versus $\protect\chi_\mathrm{cf}$ for CeCoIn$_5$,
where the $\protect\chi_\mathrm{cf}$ were obtained from bulk measurements.
\protect\cite{satoru}}
\label{fig:KKLvsChiKL}
\end{figure}

\begin{figure}[tbp]
\centering \includegraphics[width=\linewidth]{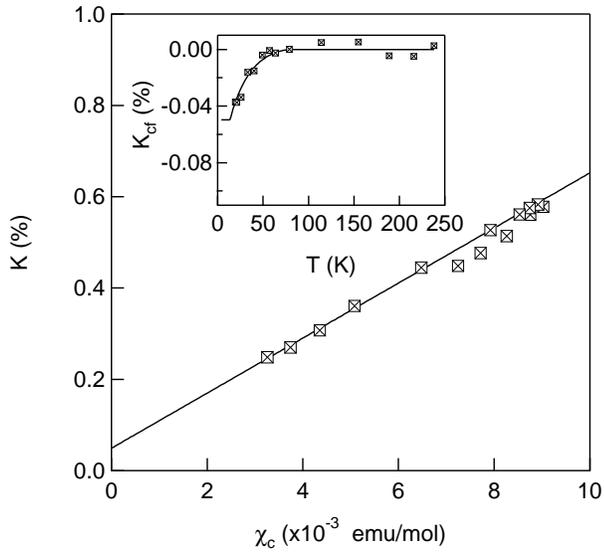}
\caption{The Si Knight shift in URu$_2$Si$_2$ versus the bulk
susceptibility. \protect\cite{bernal} The solid line is a fit to the high
temperature data. Inset: $K_\mathrm{cf}$ versus $T$, and a fit to Eq. (%
\protect\ref{eqn:KKL}).}
\label{fig:URu2Si2}
\end{figure}

\begin{figure}[tbp]
\centering \includegraphics[width=\linewidth]{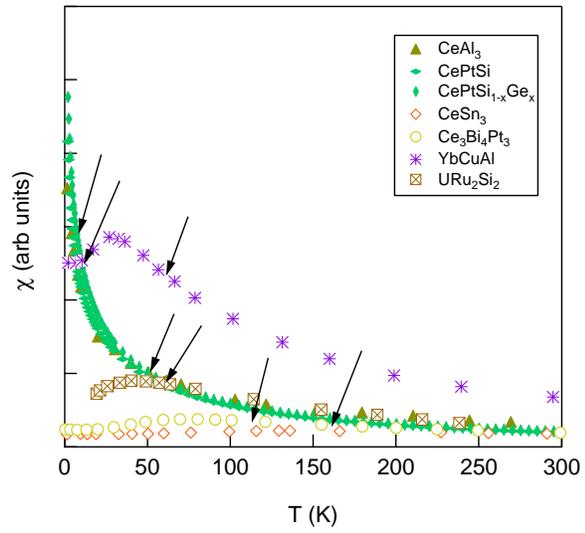}
\caption{The susceptibility $\protect\chi$ versus temperature in CeAl$_3$,
CePtSi$_{1-x}$Ge$_x$, CeSn$_3$, Ce$_3$Bi$_4$Pt$_3$, YbCuAl, and URu$_2$Si$_2$%
. $T^*$ is marked with an arrow.}
\label{fig:suscept3}
\end{figure}

Here we include Fig. \ref{fig:KKLvsChiKL}, as well as Figs. \ref%
{fig:In2CeCoIn5} through \ref{fig:URu2Si2}, showing the $K$ versus $\chi $
relationship, and the anomaly at $T^{\ast }$. The insets of the Figs. show $%
K_{\mathrm{cf}}(T)$, as well as a fit to Eq. (\ref{eqn:KKL}) . In Figs. \ref%
{fig:suscept1} - \ref{fig:suscept3} we show the bulk susceptibility versus
temperature for each compound. $T^{\ast }$ is marked with an arrow for each
material. Note that the anomalous behavior marking the emergence of the $%
\chi _{\mathrm{cf}}$ component is not obvious in the bulk susceptibility;
measurements of both the susceptibility as well as the Knight shift are
required to identify $T^{\ast }$.

\begin{figure}[h]
\centering \includegraphics[width=\linewidth]{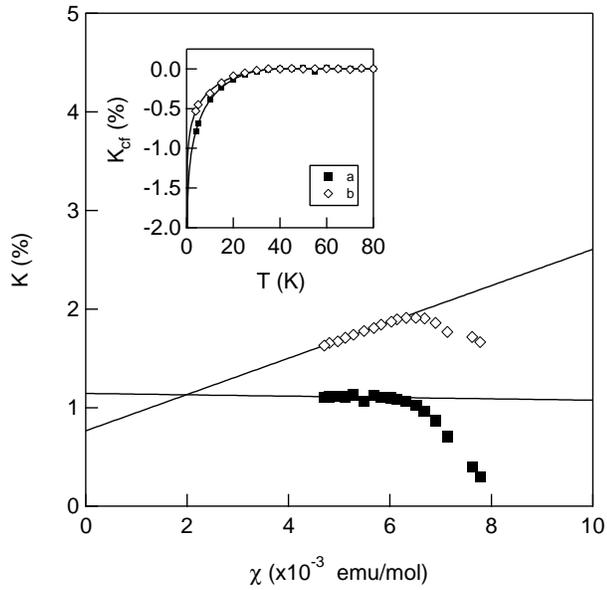}
\caption{The In(2) Knight shift CeCoIn$_5$\ versus the bulk susceptibility.%
\protect\cite{curroCeCoIn5} The solid lines are fits to the high temperature
data. Inset: $K_\mathrm{cf}$ versus $T$, and a fit to Eq. (\protect\ref%
{eqn:KKL}).}
\label{fig:In2CeCoIn5}
\end{figure}

\begin{figure}[tbp]
\centering  \includegraphics[width=\linewidth]{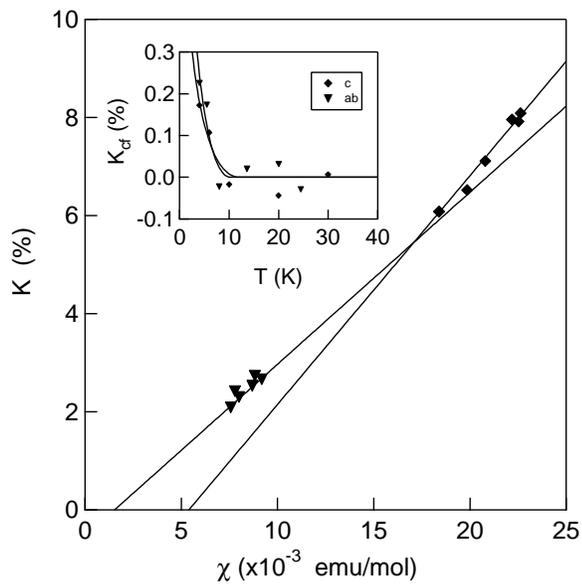}
\caption{The In(1) Knight shift in CeRhIn$_5$\ versus the bulk
susceptibility. The solid lines are fits to the high temperature data.
Inset: $K_\mathrm{cf}$ versus $T$, and a fit to Eq. (\protect\ref{eqn:KKL}).}
\label{fig:In1CeRhIn5}
\end{figure}

\begin{figure}[tbp]
\centering  \includegraphics[width=\linewidth]{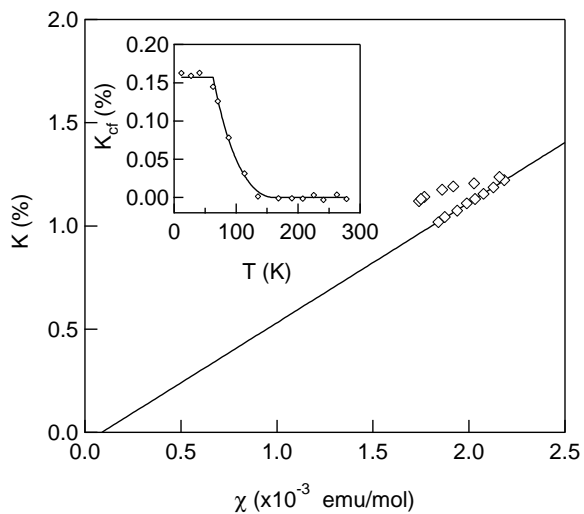}
\caption{The Sn Knight shift in CeSn$_{3}$ versus the bulk susceptibility.%
\protect\cite{malikCeSn3,CooperCeSn3} The solid lines are fits to the high
temperature data. Inset: $K_\mathrm{cf}$ versus $T$, and a fit to Eq. (%
\protect\ref{eqn:KKL}).}
\label{fig:CeSn3}
\end{figure}

\begin{figure}[tbp]
\centering  \includegraphics[width=\linewidth]{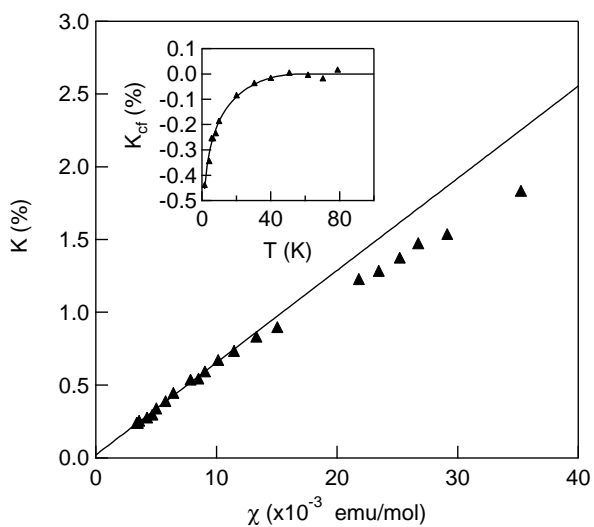}
\caption{The Al Knight shift in CeAl$_{3}$ versus the bulk susceptibility.
\protect\cite{lysakCeAl3} The solid lines are fits to the high temperature
data. Inset: $K_\mathrm{cf}$ versus $T$, and a fit to Eq. (\protect\ref%
{eqn:KKL}).}
\label{fig:CeAl3}
\end{figure}

\begin{figure}[tbp]
\centering \includegraphics[width=\linewidth]{CeCu2Si2Cuanalysis.eps}
\caption{The Cu Knight shift in CeCu$_{2}$Si$_2$ versus the bulk
susceptibility.\protect\cite{yasuokaCeCu2Si2} The solid lines are fits to
the high temperature data. Inset: $K_\mathrm{cf}$ versus $T$, and a fit to
Eq. (\protect\ref{eqn:KKL}).}
\label{fig:CuCeCu2Si2}
\end{figure}

\begin{figure}[tbp]
\centering \includegraphics[width=\linewidth]{CeCu2Si2Sianalysis.eps}
\caption{The Si Knight shift in CeCu$_{2}$Si$_2$ versus the bulk
susceptibility.\protect\cite{yasuokaCeCu2Si2} The solid lines are fits to
the high temperature data. Inset: $K_\mathrm{cf}$ versus $T$, and a fit to
Eq. (\protect\ref{eqn:KKL}).}
\label{fig:SiCeCu2Si2}
\end{figure}

\begin{figure}[tbp]
\centering  \includegraphics[width=\linewidth]{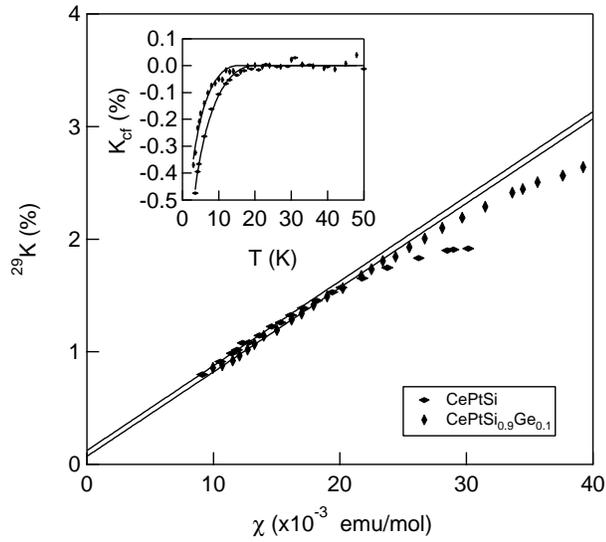}
\caption{The Si Knight shift in CePtSi$_{1-x}$Ge$_x$ for $x=0.0$ and $x=0.1$
versus the bulk susceptibility.\protect\cite{benliCePtSi} The solid lines
are fits to the high temperature data. Inset: $K_\mathrm{cf}$ versus $T$,
and a fit to Eq. (\protect\ref{eqn:KKL}).}
\label{fig:CePtSi}
\end{figure}

\begin{figure}[tbp]
\centering  \includegraphics[width=\linewidth]{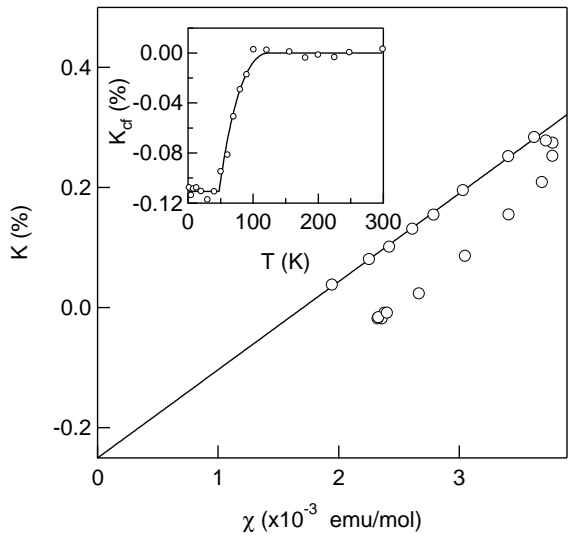}
\caption{The Bi Knight shift Ce$_3$Bi$_4$Pt$_3$ versus the bulk
susceptibility.\protect\cite{reyes343} The solid line is a fit to the high
temperature data. Inset: $K_\mathrm{cf}$ versus $T$, and a fit to Eq. (%
\protect\ref{eqn:KKL}).}
\label{fig:Ce343}
\end{figure}

\begin{figure}[tbp]
\centering  \includegraphics[width=\linewidth]{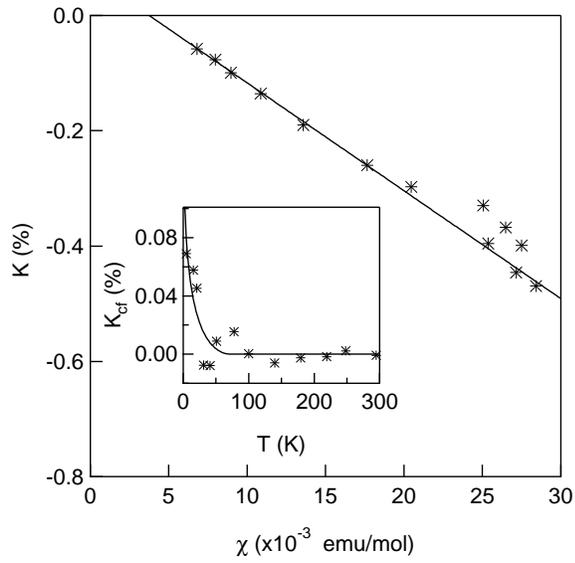}
\caption{The Cu Knight shift in YbCuAl versus the bulk susceptibility
\protect\cite{MacLaughlinYbCuAl}. The solid line is a fit to the high
temperature data. Inset: $K_\mathrm{cf}$ versus $T$, and a fit to Eq. (%
\protect\ref{eqn:KKL}).}
\label{fig:YbCuAl}
\end{figure}

\begin{figure}[tbp]
\centering  \includegraphics[width=\linewidth]{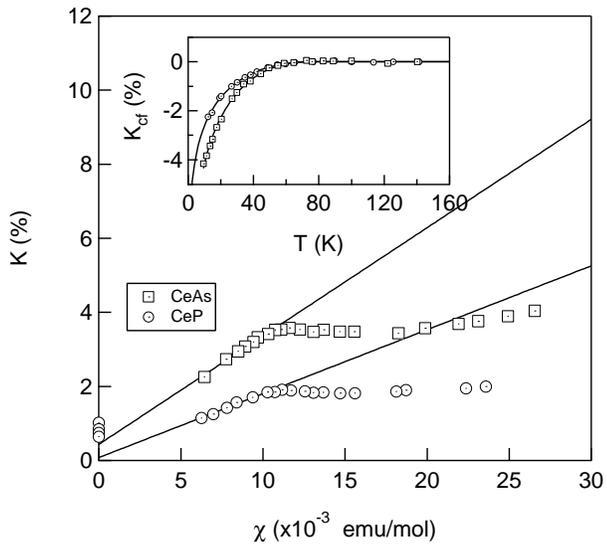}
\caption{The P and As Knight shifts in CeP and CeAs versus the bulk
susceptibility. \protect\cite{narath} The solid line is a fit to the high
temperature data. Inset: $K_\mathrm{cf}$ versus $T$, and a fit to Eq. (%
\protect\ref{eqn:KKL}).}
\label{fig:CeX}
\end{figure}

\begin{figure}[tbp]
\centering  \includegraphics[width=\linewidth]{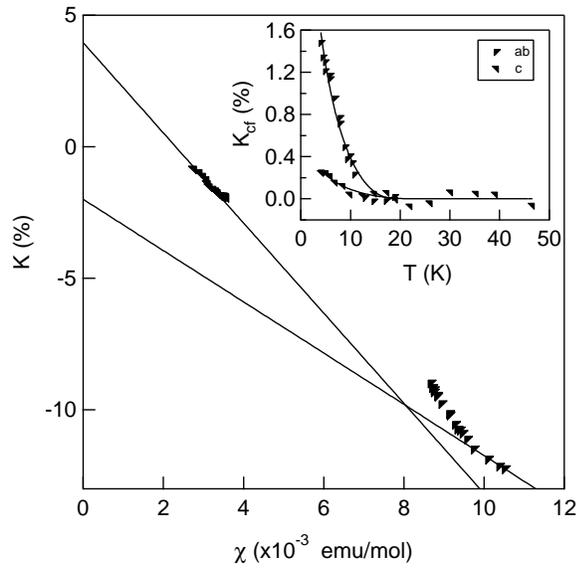}
\caption{The Pt Knight shift in UPt$_3$ versus the bulk susceptibility.
\protect\cite{deVisser,halparin} The solid line is a fit to the high
temperature data. Inset: $K_\mathrm{cf}$ versus $T$, and a fit to Eq. (%
\protect\ref{eqn:KKL}).}
\label{fig:UPt3}
\end{figure}

\begin{figure}[tbp]
\centering  \includegraphics[width=\linewidth]{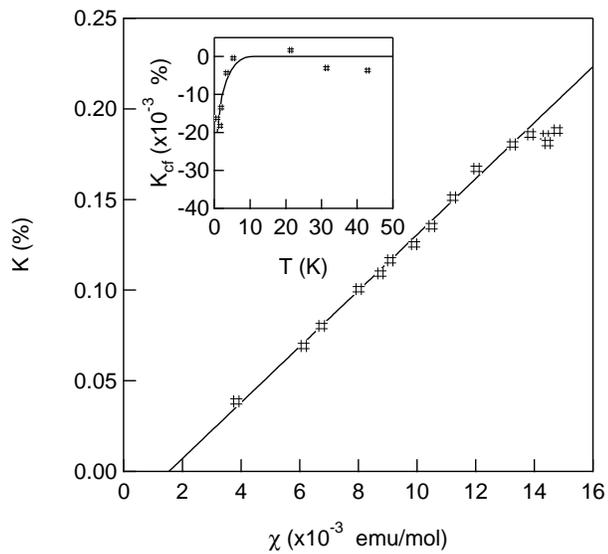}
\caption{The Pt Knight shift in UBe$_{13}$ versus the bulk susceptibility.%
\protect\cite{ClarkUBe13} The solid line is a fit to the high temperature
data. Inset: $K_\mathrm{cf}$ versus $T$, and a fit to Eq. (\protect\ref%
{eqn:KKL}).}
\label{fig:UBe13}
\end{figure}

\begin{figure}[tbp]
\centering \includegraphics[width=\linewidth]{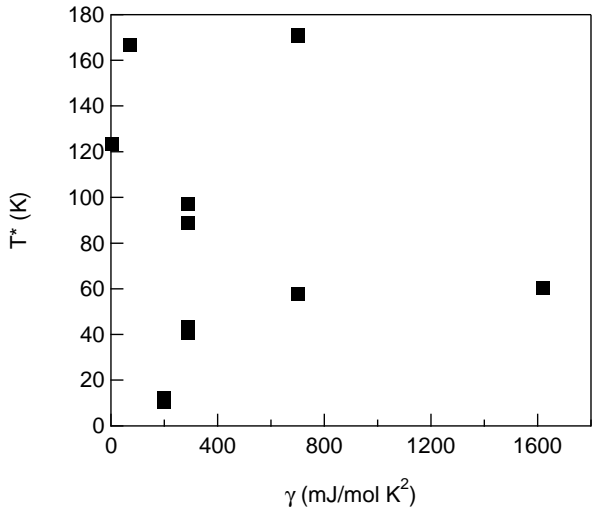}
\caption{$T^*$ versus $\protect\gamma$ for the Kondo lattice systems
discussed. There is no apparent correlation.}
\label{fig:TstarVSgamma}
\end{figure}

\begin{figure}[tbp]
\centering \includegraphics[width=\linewidth]{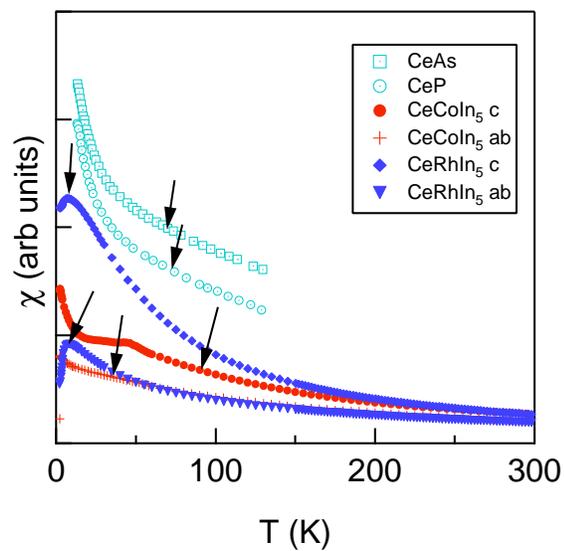}
\caption{The susceptibility $\protect\chi$ versus temperature in CeAs, CeP,
CeCoIn$_5$, and CeRhIn$_5$. $T^*$ is marked with an arrow. The data for CeAs
and CeP are offset vertically for clarity.}
\label{fig:suscept1}
\end{figure}

\begin{figure}[tbp]
\centering \includegraphics[width=\linewidth]{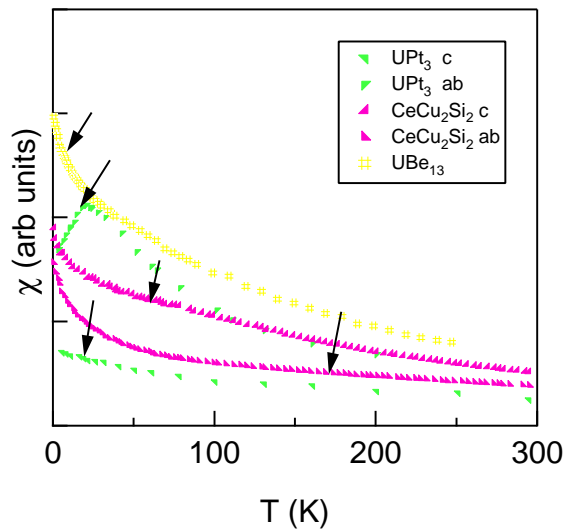}
\caption{The susceptibility $\protect\chi$ versus temperature in UPt$_3$,
CeCu$_2$Si$_2$, and UBe$_{13}$. $T^*$ is marked with an arrow.}
\label{fig:suscept2}
\end{figure}

\end{document}